\documentclass[final,12pt]{elsarticle}

\usepackage{amssymb}
\usepackage{amsmath}
\usepackage{subcaption}
\usepackage{enumitem}
\usepackage{xspace}
\usepackage{booktabs}
\usepackage{multirow}
\usepackage[normalem]{ulem}
\usepackage{xurl}
\usepackage{pifont}
\usepackage{xcolor}
\usepackage{graphicx}
\usepackage{tikz}
\usetikzlibrary{tikzmark}
\usetikzlibrary{patterns}
\usepackage{pgfgantt}
\usepackage{pgfplots}
\pgfplotsset{compat=1.18}
\usepackage{placeins}
\usepackage{float}

\newcommand{\BfPara}[1]{{\vspace{1.5mm}\noindent\bfseries #1.}\xspace}
\newcommand{\etal}{{\em et al.}\xspace}

\begin{document}

\begin{frontmatter}

\title{Adversarial Vulnerability Under Temporal Concept Drift: A Longitudinal Study of Android Malware Detection}

%% Authors
\author[1]{Ahmed Sabbah}
\author[1]{Mohammed F. Kharma}
\author[1]{Radi Jarrar}
\author[1]{Samer Zein}
\author[2]{David~Mohaisen\corref{cor1}}
\cortext[cor1]{Corresponding author}
\ead{mohaisen@ucf.edu}

%% Affiliations
\affiliation[1]{organization={Department of Computer Science, Birzeit University},
            addressline={P.O. Box 14},
            city={Birzeit},
            postcode={P627},
            state={West Bank},
            country={Palestine}}

\affiliation[2]{organization={Department of Computer Science, University of Central Florida},
            addressline={4328 Scorpius St., Building 116, Room 246},
            city={Orlando},
            postcode={32816},
            state={FL},
            country={USA}}

\begin{abstract}
As intelligent malware detection systems are deployed in evolving mobile ecosystems, both benign and malicious applications change in structure and behavior, inducing non-stationary data distributions that challenge long-term model reliability. In parallel, adversarial perturbations expose additional vulnerabilities in machine learning--based detectors. Despite extensive work on concept drift and adversarial robustness independently, their interaction in adaptive intelligent systems remains insufficiently characterized.

We present a longitudinal, drift-aware evaluation of adversarial robustness across more than a decade of Android applications using static and dynamic feature representations extracted from emulator and real-device executions. The dataset is organized into yearly slices and evaluated under three deployment protocols that emulate realistic learning scenarios: (1) same-year training and testing, (2) cross-year deployment without model updates, and (3) expanding-window retraining with cumulative historical data. Across multiple classifier families, adversarial examples are generated using FGSM and SPSA under feasibility constraints. We measure clean performance, Adversarial Accuracy (AA), Attack Success Rate (ASR), and introduce temporal linkage metrics---RobustDrop, $\Delta$ASR, and Adversarial Amplification Factor (AAF)---to quantify the relationship between distribution shift and robustness degradation.nResults show that temporal separation is associated with reduced adversarial robustness under the evaluated transfer-based feature-space setting. As the train--test gap increases, clean accuracy and adversarial accuracy decline, while attack success exhibits configuration-dependent increases, particularly under FGSM perturbations and static features. Expanding-window retraining mitigates, but does not eliminate, robustness loss under continued distributional evolution. These findings indicate that temporal drift should be considered when assessing the long-term robustness of intelligent detection systems under evolving data distributions and highlight the need for drift-aware robustness assessment frameworks in long-lived adversarial environments.

\end{abstract}


\end{frontmatter}

\section{Introduction}\label{sec:introduction}

The deployment of machine learning--based malware detectors in real-world environments exposes them to evolving data distributions, violating the stationarity assumption underlying conventional offline supervised learning~\cite{AbusnainaASAJSM25,abs-2507-22231,abs-2507-22772,AbusnainaASAJSM24,MohaisenAM15}. Intelligent decision systems operating on security data streams must contend with inherent non-stationarity, including concept drift, where the conditional relationship between inputs and labels changes as new threats, platforms, and user behaviors emerge~\cite{Moreno-TorresRACH12,GamaZBPB14}. When such drift is not explicitly monitored and accommodated, models that initially achieve strong performance may degrade over time, increasing missed detections and undermining the reliability of long-lived intelligent systems.

Android provides a representative and operationally relevant setting for studying learning under distribution shift. As the dominant mobile operating system and a primary malware target~\cite{ShenVMKZ19,AlasmaryKAPCAAN19,ShenVMKZ19,MobileOp68:online,MalwareS28:online,Kaspersk70:online}, it has motivated extensive research on data-driven detection using static, dynamic, and hybrid feature representations across both traditional and deep learning architectures~\cite{PanGFF20,Alzubaidi21,LiFWCZYWG22}. Although many studies report high accuracy on curated benchmarks, most evaluations implicitly assume that training and test samples are drawn from the same temporal distribution. This assumption limits insight into long-term system behavior and the robustness of intelligent detectors deployed in evolving environments.

Recent work has begun to revisit this stationarity assumption by examining temporal effects in Android malware detection and related domains. Concept drift has been characterized at both dataset and behavioral levels using time-indexed collections such as KronoDroid and other longitudinal datasets~\cite{Guerra-ManzanaresBN21,Guerra-ManzanaresLB22,Guerra-ManzanaresB22a}. These studies demonstrate that performance can decline when models are evaluated on future time windows or across changing environments, and that improper temporal partitioning may introduce leakage and optimistic bias~\cite{PendleburyPJKC19,BarberoPPC22}. However, prior analyses largely emphasize clean-data performance and mitigation strategies, leaving the relationship between temporal distribution shift and adversarial vulnerability insufficiently explored from the perspective of robust intelligent systems.

In parallel, adversarial machine learning has shown that small, carefully crafted perturbations can induce misclassification across a wide range of modern models~\cite{GoodfellowSS15,MadryMSTV18,CarliniW17}. Within malware detection, classifiers can be evaded through both feature-space and problem-space manipulations~\cite{PierazziPCC20,BostaniM24}. Yet most adversarial evaluations assume a fixed underlying data distribution and do not account for how vulnerability profiles may evolve as the data-generating process shifts over time.

Consequently, two lines of research have progressed largely independently: \ding{172} the study of temporal degradation in malware detectors under concept drift and \ding{173} the analysis of adversarial evasion under static training conditions. Their interaction in time-evolving deployment scenarios remains insufficiently characterized. In particular, it is unclear whether increasing temporal separation systematically amplifies adversarial vulnerability beyond its impact on clean performance, and whether incremental retraining strategies that partially address drift also improve robustness or reshape the effective attack surface.

In this study, we investigate this interaction using static and dynamic feature representations extracted from a temporally ordered Android dataset spanning multiple years, including executions on both emulators and real devices. We evaluate two deployment protocols that emulate realistic learning scenarios: (i) a cross-year setting, where a model trained on a single year is evaluated on subsequent years without updates, and (ii) an expanding-window setting, where the training set grows cumulatively over time to incorporate newly observed data. Our contributions are as follows.

\begin{itemize}[leftmargin=*]

  \item We present a longitudinal evaluation of intelligent malware detection systems under non-stationary data distributions. Using a time-ordered dataset with static and dynamic feature representations, we systematically quantify how predictive performance and adversarial robustness evolve as the temporal gap between training and evaluation increases under cross-year and expanding-window learning protocols.

  \item We conduct a drift-aware adversarial robustness analysis using transfer-based feature-space attacks. Surrogate models are trained on temporally consistent splits, and adversarial examples are generated via FGSM and SPSA under explicit feasibility constraints. Robustness is assessed across diverse classifier families, including RF, GB, KNN, CNN, RNN, LSTM, and GRU, to capture heterogeneous learning behaviors.

  \item We define and use drift--robustness linkage metrics, RobustDrop, $\Delta\mathrm{ASR}$, and the Adversarial Amplification Factor (AAF), as temporal contrast measures relative to the same-year baseline. Unlike reporting AA and ASR alone for each train--test pair, these metrics make robustness loss comparable across training years, feature spaces, attacks, and deployment protocols.

  \item We comparatively analyze static deployment (cross-year) and adaptive retraining (expanding-window) strategies. While cumulative retraining mitigates degradation in both clean and adversarial performance, it does not fully restore same-year robustness, highlighting persistent reliability challenges in long-lived intelligent detection systems.

\end{itemize}

\BfPara{Organization}
Section~\ref{sec:Motivation} outlines the motivation and formalizes the research questions in the context of learning under non-stationary and adversarial conditions. Section~\ref{DriftSection} reviews concept drift and its underlying causes, with emphasis on distributional evolution in adaptive intelligent systems. Section~\ref{sec:related} surveys prior work on temporal robustness and adversarial vulnerability. Section~\ref{sec:method} describes the experimental methodology, including the longitudinal dataset construction, feature representations, temporal evaluation protocols, learning models, adversarial generation procedures, and robustness metrics. Section~\ref{sec:results} presents the empirical findings and analyzes performance trends across temporal and adversarial settings. Section~\ref{sec:Threats} discusses threats to validity and experimental limitations. Section~\ref{sec:conclusion} concludes the paper and outlines implications for drift-aware robustness evaluation in evolving intelligent systems.

\section{Motivation and Research Questions}\label{sec:Motivation}

Intelligent malware detection systems deployed in real-world environments operate under non-stationary conditions, where evolving application behavior induces temporal distribution shift. Such non-stationarity challenges the reliability of learned decision boundaries over time. While prior work has examined concept drift and adversarial evasion independently, their interaction in adaptive, time-evolving deployment scenarios remains insufficiently characterized from a systems perspective.

\BfPara{Research Questions}
\begin{itemize}[leftmargin=*]
  \item \textbf{RQ1:} How does temporal concept drift affect predictive performance as the train--test year gap increases in longitudinal learning settings?

  \item \textbf{RQ2:} How does adversarial vulnerability evolve under increasing temporal separation between training and evaluation data?

  \item \textbf{RQ3:} What quantitative relationship exists between drift-induced performance degradation and adversarial vulnerability, and to what extent can expanding-window retraining mitigate this interaction in adaptive learning systems?
\end{itemize}

\paragraph{Temporal Evaluation}

To investigate these questions, we design three complementary deployment protocols that emulate realistic intelligent system lifecycles:

\begin{enumerate}[leftmargin=*]
  \item[\ding{172}] \textbf{Same-year.} Train on year $Y$ and evaluate on held-out data from $Y$, establishing an in-distribution reference for both predictive accuracy and adversarial robustness.

  \item[\ding{173}] \textbf{Cross-year.} Train on year $Y_1$ and evaluate on future years $Y_2 > Y_1$, exposing a fixed model to distributional evolution without adaptation.

  \item[\ding{174}] \textbf{Expanding-window.} Train cumulatively on data up to year $Y_i$ and evaluate on $Y_{i+1}$, emulating periodic retraining with all historically available observations.
\end{enumerate}

For each protocol, adversarial examples are generated using FGSM and SPSA under explicit feature-space feasibility constraints. We report predictive accuracy, Adversarial Accuracy (AA), and Attack Success Rate (ASR), and employ temporal linkage metrics (RobustDrop, $\Delta$ASR, AAF) to systematically quantify the interaction between distributional shift and adversarial vulnerability in evolving learning environments.

\section{Concept Drift}\label{DriftSection}

Concept drift refers to temporal changes in the joint distribution $P(x,y)$ of features $x$ and labels $y$~\cite{SchlimmerG86}. In non-stationary learning environments, such distributional evolution can substantially degrade predictive performance when models trained on historical data are deployed on future observations. From the perspective of intelligent systems, concept drift challenges the stability of learned decision boundaries and necessitates evaluation frameworks that account for time-dependent data generation processes.

In Android malware detection, drift may arise from multiple sources, including API evolution, changes in application development practices, behavioral shifts in benign applications, and the emergence, mutation, or disappearance of malware families. These factors induce structural and behavioral changes in feature distributions over time, affecting both classification performance and robustness properties.

%R7A5. R7A9, R7A3
\BfPara{Connection to adversarial settings}
In this work, we distinguish natural temporal drift from test-time adversarial perturbations. Natural drift arises from ecosystem changes over time, such as platform evolution, benign application behavior, and malware-family changes. Adversarial perturbations do not create drift in our longitudinal dataset, instead, they test whether models trained on earlier distributions become more sensitive to constrained feature changes when evaluated on future data. Thus, our study examines how adversarial vulnerability changes under naturally evolving temporal distributions, rather than claiming that individual adversarial examples cause real drift.

\BfPara{Root Causes} Following Xiang \etal~\cite{XiangZCW23}, drift can be categorized into three principal forms:

\begin{itemize}[leftmargin=*]

    \item \textbf{Virtual drift:} $P_{t_0}(x) \neq P_{t_1}(x)$ while $P_{t_0}(y \mid x) = P_{t_1}(y \mid x)$. Here, the marginal feature distribution shifts over time, but the underlying conditional decision boundary remains stable. In such cases, performance degradation may result from representation shift rather than changes in class semantics.

    \item \textbf{Real drift:} $P_{t_0}(x) = P_{t_1}(x)$ while $P_{t_0}(y \mid x) \neq P_{t_1}(y \mid x)$. The conditional relationship between features and labels changes, indicating that previously learned decision boundaries are no longer valid.

    \item \textbf{Hybrid drift:} Both $P(x)$ and $P(y \mid x)$ change simultaneously. This scenario is common in dynamic and adversarial environments, where structural feature evolution and shifting class semantics co-occur. The longitudinal setting considered in this work captures such combined effects within the KronoDroid timeline.

\end{itemize}

Understanding these drift mechanisms is essential for analyzing how predictive performance and adversarial robustness evolve in long-lived intelligent detection systems operating under temporal non-stationarity.

\section{Related Work}\label{sec:related}

\subsection{Malware Detection and Concept Drift}

Supervised learning methods are commonly developed under the assumption of stationary data distributions. Concept drift, defined as temporal changes in the relationship between inputs and targets~\cite{GamaZBPB14}, challenges this assumption and has motivated research on learning under non-stationarity~\cite{CeschinBGPOG23,TripathiGB25}. In Android malware detection, numerous studies demonstrate the effectiveness of machine learning approaches while recognizing the impact of temporal evolution on predictive performance~\cite{HuMZLYL17}. Chen \etal~\cite{ChenZKYCPPCW23} analyzed feature-space and data-space drift and quantified their respective effects on Android and PE malware detectors across heterogeneous feature representations.

Guerra-Manzanares \etal~\cite{Guerra-Manzanares23} examined drift using dynamic system-call features and emphasized the importance of timestamp-aware modeling for capturing temporal evolution. They later proposed a classifier-pool framework for temporally ordered datasets, showing that time-aware training strategies influence detection accuracy~\cite{Guerra-ManzanaresB22a}. Their adaptive ensemble selection highlights the need for dynamic model management in evolving environments.

Chow \etal~\cite{ChowKLCAP23} investigated the structural causes of drift and attributed performance degradation to the emergence of new malware families and the evolution of existing ones. Complementary work has shown that adversarial manipulation can further exacerbate degradation in drifted settings~\cite{AbusnainaWAWCM23}. Abusnaina \etal~\cite{AbusnainaAAAJNM22} evaluated robustness under temporal variation and reported substantial accuracy reductions in non-stationary environments.

Ceschin \etal~\cite{CeschinBGPOG23} studied temporal drift using the DREBIN and AndroZoo datasets over a nine-year period. Employing Word2Vec and TF-IDF representations with adaptive random forest and stochastic gradient descent classifiers, they demonstrated that evolving data distributions necessitate continual model updates. Their adaptive strategy improved F1 scores by $22.05\%$ on DREBIN and $8.77\%$ on AndroZoo, underscoring the importance of adaptive learning in longitudinal malware detection.

\subsection{Adversarial Concept Drift and Robustness}

In deployed intelligent detection systems, adversaries may deliberately adapt their strategies to evade learned models, inducing distributional shifts referred to as \emph{adversarial concept drift}~\cite{HinderVH24}. Unlike drift caused by natural ecosystem evolution, adversarial drift results from strategic manipulation intended to degrade model performance.

In Android malware detection, such adaptations include obfuscation, polymorphism, and metamorphism~\cite{FarukiBJBMP23}, which alter observable artifacts while preserving malicious functionality. Adversarial machine learning further enables evasion through carefully crafted perturbations that exploit model vulnerabilities~\cite{SethiK18,KoryckiK23}.

Adversarial attack models are typically categorized by the adversary’s knowledge. In white-box settings, access to model parameters allows gradient-based methods such as FGSM~\cite{GoodfellowSS15}, PGD~\cite{MadryMSTV18}, and optimization-based attacks~\cite{CarliniW17}. In black-box scenarios, attackers exploit transferability or query-efficient approaches such as ZOO~\cite{ChenZSYH17} and related adaptations for malware classifiers~\cite{RosenbergSER20}. Gray-box attacks assume partial knowledge and may involve problem-space manipulations tailored to malware detection~\cite{PierazziPCC20}.

Although many adversarial methods were originally developed for continuous domains, they have been adapted to discrete malware representations, including API calls, permissions, and opcode sequences. Functionality-preserving strategies include genetic search~\cite{YustePT22}, reinforcement learning–based modification~\cite{AndersonKFER18}, and GAN-based generation~\cite{Hu022}. In Android settings, these techniques enable benign code insertion, instruction reordering, or manifest modification to alter feature distributions while maintaining executable behavior~\cite{BostaniM24}.

Collectively, existing research has examined temporal degradation and adversarial robustness largely in isolation. Recent studies have begun to examine their intersection in related security domains, such as adversarial perturbations under concept drift in intrusion detection~\cite{ApruzzeseFP24}. However, a unified longitudinal evaluation that explicitly quantifies this interaction in Android malware detection remains limited in Android malware detection, where temporal drift, feature-space adversarial robustness, and retraining protocols are rarely evaluated together over long time spans. Our study addresses this gap by evaluating clean and adversarial performance across same-year, cross-year, expanding-window, size-matched expanding-window, and recent window settings using static and dynamic Android malware features.

\section{Methodology}\label{sec:method}

Figure~\ref{fig:pipeline} summarizes the longitudinal evaluation framework, including data collection, feature extraction, yearly partitioning, temporal learning protocols, surrogate-based adversarial generation, and drift-aware robustness metrics. For consistency with the notation introduced in Section~\ref{sec:introduction}, we use AA for Adversarial Accuracy, ASR for Attack Success Rate, RobustDrop for adversarial-accuracy reduction relative to the same-year baseline, $\Delta\mathrm{ASR}$ for the corresponding ASR change, and AAF for the Adversarial Amplification Factor.

\subsection{Dataset Overview}\label{sec:DatasetOverview}

\begin{figure}[H]
    \centering
    \begin{tikzpicture}
        \begin{axis}[
            ybar,
            width=14cm,
            height=4.2cm,
            ylabel={\# Apps},
            symbolic x coords={2008, 2009, 2010, 2011, 2012, 2013, 2014, 2015, 2016, 2017, 2018, 2019, 2020},
            xtick=data,
            x tick label style={rotate=45, anchor=east},
            ymin=0,
            bar width=5pt,
            ymajorgrids=true,
            enlargelimits=false,
            tick label style={font=\scriptsize},
            legend style={at={(0.75,1.2)}, anchor=north, legend columns=2, font=\tiny},
            nodes near coords,
            every node near coord/.append style={font=\tiny, rotate=90, anchor=west},
            enlarge x limits=0.05
        ]

        % Benign - Real Device
        \addplot[ybar, draw=blue!80!black, fill=blue!35, pattern=north east lines, pattern color=blue!80!black] coordinates {
            (2008, 64) (2009, 622) (2010, 5074) (2011, 22873)
            (2012, 342) (2013, 489) (2014, 632) (2015, 720)
            (2016, 743) (2017, 650) (2018, 775) (2019, 200) (2020, 275)
        };
        \addlegendentry{Benign - Real Device}

        % Malware - Real Device
        \addplot[ybar, draw=red!80!black, fill=red!35, pattern=north west lines, pattern color=red!80!black] coordinates {
            (2008, 934) (2009, 20) (2010, 269) (2011, 3137)
            (2012, 7564) (2013, 7487) (2014, 8005) (2015, 1424)
            (2016, 2445) (2017, 4806) (2018, 4006) (2019, 1491) (2020, 1048)
        };
        \addlegendentry{Malware - Real Device}

        % Benign - Emulator
        \addplot[ybar, draw=teal!80!black, fill=teal!35, pattern=grid, pattern color=teal!80!black] coordinates {
            (2008, 66) (2009, 607) (2010, 4978) (2011, 22146)
            (2012, 309) (2013, 444) (2014, 596) (2015, 689)
            (2016, 715) (2017, 788) (2018, 733) (2019, 180) (2020, 1002)
        };
        \addlegendentry{Benign - Emulator}

        % Malware - Emulator
        \addplot[ybar, draw=orange!90!black, fill=orange!40, pattern=crosshatch, pattern color=orange!90!black] coordinates {
            (2008, 686) (2009, 19) (2010, 260) (2011, 2277)
            (2012, 6498) (2013, 5641) (2014, 6286) (2015, 1031)
            (2016, 1941) (2017, 625) (2018, 2299) (2019, 1388) (2020, 256)
        };
        \addlegendentry{Malware - Emulator}

        \end{axis}
    \end{tikzpicture}
    \vspace{-3mm}
    \caption{Yearly distribution of benign and malware samples in the KronoDroid real-device and emulator partitions from 2008 to 2020.}
    \label{fig:Kdataset}
    \vspace{-3mm}
\end{figure}
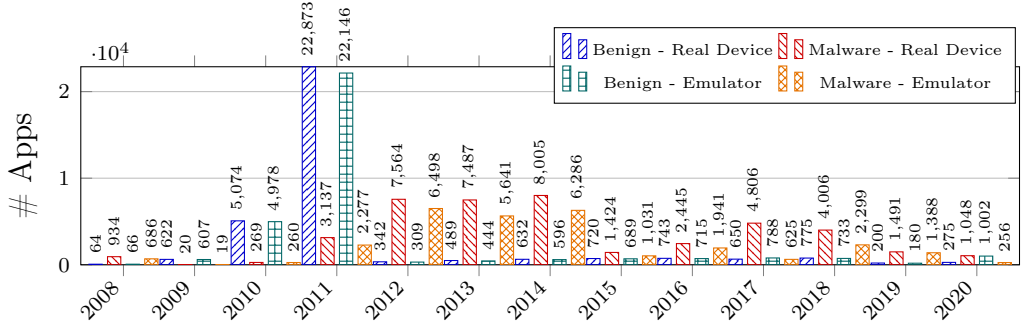

\begin{figure}[H]
    \centering
    \includegraphics[width=\linewidth]{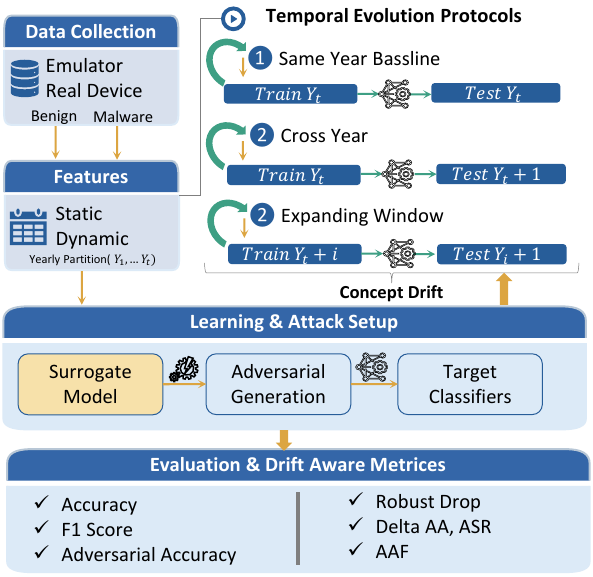}
    \caption{Temporal drift evaluation pipeline. Applications are executed on emulators and real devices, features are extracted and grouped by year, models are trained under same-year, cross-year, and expanding-window protocols, and evaluated under FGSM and SPSA attacks.}
    \label{fig:pipeline}
\end{figure}

We use the KronoDroid dataset~\cite{Guerra-ManzanaresBN21}, which provides temporally indexed annual slices spanning 2008 to 2020. The dataset includes static features (permissions and intents) and dynamic features (system call counts) for malware and benign applications collected from both emulators and real devices.
The real-device partition contains 36,755 benign samples and 41,382 malware samples, while the emulator partition contains 35,246 benign samples and 28,745 malware samples. Figure~\ref{fig:Kdataset} shows the yearly class distribution for both partitions. Although the aggregate class balance is moderate, the yearly slices are not uniformly balanced, with noticeable variation in sample counts across years, motivating the use of stratified same-year splits and macro-averaged evaluation metrics.

To enable longitudinal analysis under non-stationarity, we construct per-year subsets based on each application's last update timestamp, ensuring temporally consistent evaluation splits.

\BfPara{Feature spaces}
We evaluate both static and dynamic feature representations, static, including binary permission and intent indicators extracted from the application package, and dynamic features, which are system-call count vectors collected during execution. In our processed data, each partition contains 166 static features and 288 dynamic features. Static features mainly reflect declared application structure and Android API usage, while dynamic features capture runtime behavior and may be more sensitive to execution environment and malware-family evolution over time.

\begin{itemize}[leftmargin=*]

    \item \textbf{Validity projection.} Dynamic features are constrained to non-negative integers and static features to binary values. After each adversarial update, feature vectors are projected back to the feasible domain to eliminate invalid values.

    \item \textbf{Selective standardization.} Only dynamic features are standardized using statistics computed from the corresponding training-year split. This stabilizes gradient-based perturbations while preventing temporal leakage. Static features remain unscaled. Perturbations are generated in standardized space and inverted prior to evaluation.

    \item \textbf{Bounded edit budgets.} Each attack modifies at most $k_{\text{static}}=\lceil 0.05\, d_{\text{static}}\rceil$ and $k_{\text{dynamic}}=\lceil 0.05\, d_{\text{dynamic}}\rceil$ features. Dynamic updates are further restricted to at most $\pm 1$ per coordinate. These constraints maintain validity in the feature representation while enabling a controlled robustness evaluation.

\end{itemize}

These design choices ensure that adversarial samples remain within the feasible representation space, enabling systematic robustness assessment under temporal distribution shift.

\subsection{Models and Surrogate}

To examine heterogeneous learning behavior under drift, we evaluate seven target models: RF, GB, KNN, CNN, RNN, GRU, and LSTM. Because several targets are non-differentiable or exhibit unstable gradients, we train a logistic regression surrogate on each temporally consistent training split and generate adversarial examples for transfer-based evaluation.

The surrogate consists of a single linear layer optimized with Adam (learning rate $10^{-2}$) for 20 epochs, providing stable gradients in sparse feature spaces. Standardization parameters are computed exclusively from the corresponding training-year split to avoid temporal leakage. After adversarial generation, standardized perturbations are inverted and projected to the valid domain prior to evaluation.

\subsection{Adversarial Generation}
Transfer-based surrogate attacks are commonly used to study adversarial transferability under black-box evaluation settings, including evaluations involving deep neural networks~\cite{PapernotMGJCS17,LiuCLS17}. In malware detection, feature-space adversarial examples have also been studied for neural malware classifiers~\cite{GrossePMBM17}. However, we do not assume that the logistic regression surrogate accurately approximates the target decision boundary.  Instead, it provides a common differentiable reference model for generating constrained feature-space perturbations, which are then evaluated by transferring them to the target classifiers. Accordingly, the reported results should be interpreted as transfer-based feature-space robustness observations rather than direct estimates of intrinsic target-model robustness.

We use FGSM and SPSA because they represent two complementary perturbation-generation mechanisms under the same constrained feature-space setting. The FGSM is a first-order gradient-based method that provides an efficient way to test the sensitivity to gradient-aligned perturbations. SPSA is a zeroth-order finite-difference method that estimates the perturbation directions without requiring analytical gradients. In our experiments, both attacks are generated on the logistic regression surrogate and then transferred to the target classifiers. Therefore, the evaluation should be interpreted as a transfer-based feature-space attack setting rather than a white-box attack on the target models. In particular, SPSA is not used as a direct black-box attack against the target classifiers because the target models are not queried during adversarial optimization.

\BfPara{FGSM} We apply FGSM as a surrogate-gradient attack~\cite{GoodfellowSS15}. For each input, we compute the gradient of the surrogate loss and take a single step of size $\varepsilon$ in the sign direction. Following prior malware studies~\cite{PierazziPCC20,AndersonKFER18}, we use $\varepsilon=0.55$ for static features and $\varepsilon=0.20$ for dynamic features. Edits are restricted to the top-$k$ gradient coordinates within each feature block and projected to the feasible domain through rounding, clamping, and bounded per-coordinate updates.

We use SPSA as a surrogate-based zeroth-order perturbation method. SPSA estimates finite-difference directions on the surrogate model and is then evaluated by transfer to the target classifiers. Thus, in this study, SPSA should not be interpreted as a direct query-based black-box attack on the target models.
\BfPara{SPSA} We employ SPSA~\cite{UesatoOKO18}, a zeroth-order optimization method adapted to discrete malware representations~\cite{RosenbergSER20}. SPSA estimates finite-difference directions on the surrogate model and is then evaluated by transfer to the target classifiers. Thus, in this study, SPSA should not be interpreted as a direct query-based black-box attack on the target models. Gradients are estimated using two-sided random perturbations with scale $\delta=0.01$. We perform 50 optimization steps with three random restarts and apply the final perturbation with step size $\epsilon=0.20$. As with the FGSM, updates are restricted to blockwise top-$k$ coordinates and projected to the feasible domain.

\BfPara{Projection Diagnostics}
In this work, feasibility refers to representation-level validity rather than problem-space APK validity. Static features must remain binary, dynamic features must remain non-negative integer counts, and perturbations must respect the imposed editing budget. Therefore, adversarial examples are projected back to the feasible feature domain after generation.
Because this projection can weaken or remove perturbations, a low ASR should not be interpreted automatically as strong model robustness. To make this effect explicit, we report the perturbation diagnostics before and after the projection. Specifically, we measure the raw $L_0$ perturbation size before projection, the projected $L_0$ perturbation size after enforcing feature-domain constraints, and the projection-removal percentage, which is defined as the fraction of raw perturbation coordinates removed by the projection step. These diagnostics show whether an attack remains active after projection or whether feasible domain mapping removes most perturbations. These constraints enforce representation-level validity in the model. They do not guarantee that every perturbed feature vector corresponds to a functional, installable, or behavior-preserving APK. Accordingly, $\Delta\mathrm{ASR}$ and AAF are not interpreted for SPSA configurations where ASR is near-zero or constant; SPSA is retained only as an auxiliary constrained diagnostic in those cases.

\subsection{Temporal Learning Protocols}\label{sec:TemporalProtocols}

To emulate realistic intelligent system lifecycles under non-stationarity, we design three complementary evaluation protocols.

\BfPara{Same-year} For each year $Y$, we construct an in-distribution baseline by splitting data into $70\%$ training and $30\%$ stratified holdout sets. This baseline defines reference performance under stationary conditions.

\BfPara{Cross-year} We train on a single year $Y_1$ and evaluate on all years $Y_2 \ge Y_1$. When $Y_2 > Y_1$, performance reflects temporal generalization without model adaptation.

\BfPara{Expanding-window} We train cumulatively on data from $Y_{\min}$ through $Y_1$ and evaluate on $Y_2 > Y_1$. This protocol emulates periodic retraining with all historically available data, reflecting the adaptive learning in evolving environments.

\BfPara{Additional control baselines}
In addition to the original expanding-window protocol, we add two control baselines. The size-matched expanding-window protocol samples from the cumulative historical pool up to year $Y_1$ while matching the training size of the corresponding single-year model. This controls whether expanding-window gains mainly result from larger training sets. The recent-window protocol trains only on the most recent three years up to $Y_1$, allowing comparison between cumulative historical training and recency-focused drift adaptation.

\subsection{Evaluation Metrics}
For each $(Y_1,Y_2)$ pair, we compute predictive and robustness metrics.

\BfPara{Predictive Metrics} Accuracy, precision, recall, and macro F1 are computed on the clean $Y_2$ test set. When $Y_1=Y_2$, results correspond to the same-year baseline; when $Y_2>Y_1$, they quantify temporal generalization.

Let $\{(x_i,y_i)\}_{i=1}^{n}$ denote the clean test set of size $n$, 
where $x_i$ is a test sample with ground-truth label $y_i$. 
Let $x'_i$ denote the adversarial counterpart of $x_i$, 
$f(\cdot)$ the evaluated classifier, 
and $\mathbf{1}[\cdot]$ the indicator function.

\BfPara{Adversarial Accuracy (AA)}
Adversarial Accuracy measures the fraction of adversarial samples that remain correctly classified:

\[
AA = \frac{1}{n}\sum_{i=1}^{n}\mathbf{1}\!\left[f(x'_i)=y_i\right].
\]

\BfPara{Attack Success Rate (ASR)} Attack Success Rate quantifies the proportion of originally correct predictions that are flipped by adversarial perturbations:

\[
\text{ASR} =
\frac{\sum_{i=1}^{n}\mathbf{1}\!\left[f(x_i)=y_i \land f(x'_i)\neq y_i\right]}
{\sum_{i=1}^{n}\mathbf{1}\!\left[f(x_i)=y_i\right]}.
\]

\BfPara{Drift--Robustness Linkage Metrics} To explicitly quantify the interaction between temporal separation and adversarial vulnerability, we define:

\[
\mathrm{RobustDrop} = \mathrm{AA}_{Y_1,\text{baseline}} - \mathrm{AA}_{Y_2},
\]

\[
\Delta\mathrm{ASR} = \mathrm{ASR}_{Y_2} - \mathrm{ASR}_{Y_1,\text{baseline}},
\]

\[
\mathrm{AAF} =
\frac{\mathrm{ASR}_{Y_2}}
{\mathrm{ASR}_{Y_1,\text{baseline}} + 10^{-3}}.
\]

Here, ``baseline'' refers to the same-year holdout at $Y_1$ under the corresponding protocol. Positive $\mathrm{RobustDrop}$ and $\Delta\mathrm{ASR}$ indicate increased vulnerability under temporal separation, while $\mathrm{AAF}>1$ indicates higher ASR relative to the same-year baseline. Because AAF is ratio-based, it can be unstable when the baseline ASR is close to zero. Therefore, AAF is interpreted only when baseline ASR is non-negligible and is read jointly with RobustDrop and $\Delta\mathrm{ASR}$. The constant $10^{-3}$ prevents division by zero but does not remove this interpretability limitation.

These linkage metrics complement AA and ASR. Where AA and ASR describe adversarial performance for a specific train--test pair, whereas RobustDrop, $\Delta\mathrm{ASR}$, and AAF express how that performance changes relative to the same-year baseline. This temporal contrast makes robustness degradation easier to compare across training years, feature spaces, attacks, and deployment protocols.

\subsection{Implementation Details}
All experiments use a fixed random seed of 42 for data splitting and model initialization when supported by the implementation. RF, GB, and KNN are implemented using scikit-learn with the default settings except for fixed random states where applicable. Neural models are implemented in Keras and trained for 12 epochs with batch size 128 using binary cross-entropy and Adam. The CNN uses two Conv1D layers with 64 and 128 filters, followed by a dense layer with 128 units and dropout 0.2. The RNN, LSTM, and GRU models use 8 recurrent units with return sequences enabled, followed by a dense layer with 128 units and a dropout of 0.2. The logistic regression surrogate is trained for 20 epochs using Adam with a learning rate $10^{-2}$.

% -------------------------------------------------
\section{Results and Discussion}\label{sec:results}

We report empirical results for both emulator and real-device datasets using static and dynamic feature representations under cross-year and expanding-window protocols.

\subsection{RQ1: Clean Performance Under Temporal Drift}

\begin{figure}[H]
    \centering

    \begin{subfigure}[t]{0.99\textwidth}
        \centering
        \includegraphics[width=\textwidth]{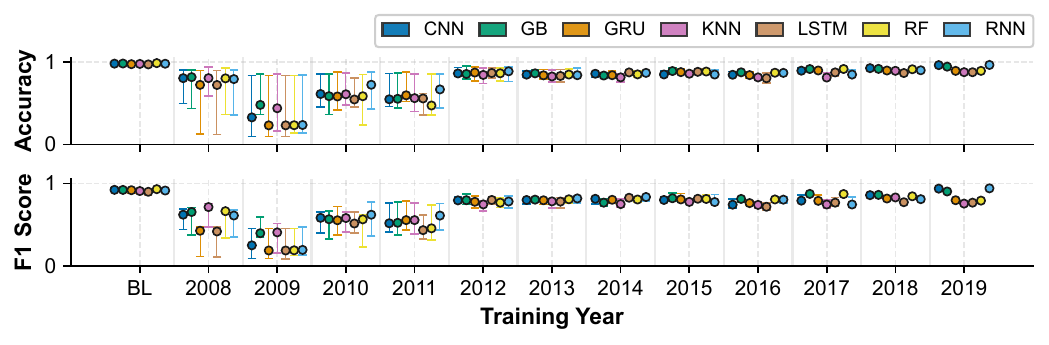}
        \caption{Real device with static features.}
        \label{fig:rq1-real-static-cross}
    \end{subfigure}

    \vspace{0.8em}

    \begin{subfigure}[t]{0.99\textwidth}
        \centering
        \includegraphics[width=\textwidth]{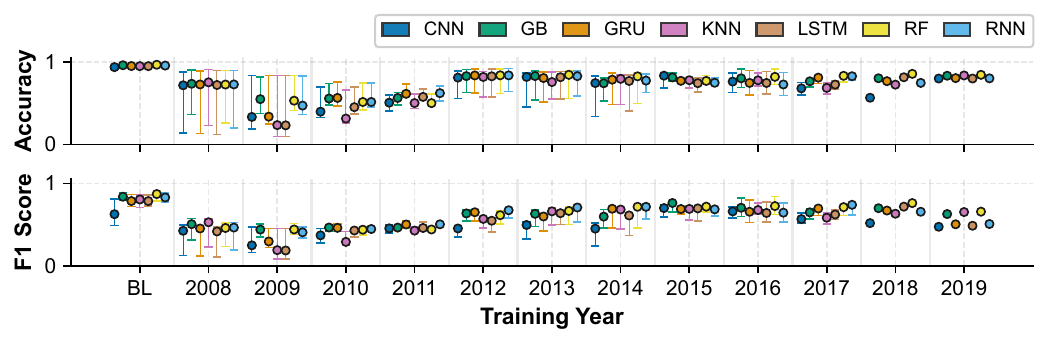}
        \caption{Real device with dynamic features.}
        \label{fig:rq1-real-dynamic-cross}
    \end{subfigure}

    \caption{Clean performance over time under the cross-year protocol for the real device dataset, using static and dynamic features. For each training year $Y_1$, point plots show the distribution of accuracy and F1 score over all test years $Y_2 > Y_1$ and all classifiers.}
    \label{fig:rq1-cross}
\end{figure}

\begin{figure}[H]
    \centering

    \begin{subfigure}[t]{0.99\textwidth}
        \centering
        \includegraphics[width=\textwidth]{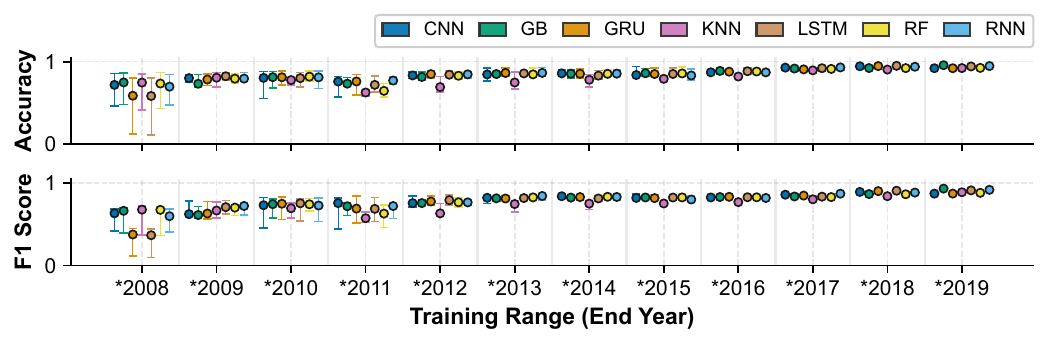}
        \caption{Emulator with static features.}
        \label{fig:rq1-emu-static-inc}
    \end{subfigure}

    \vspace{0.8em}

    \begin{subfigure}[t]{0.99\textwidth}
        \centering
        \includegraphics[width=\textwidth]{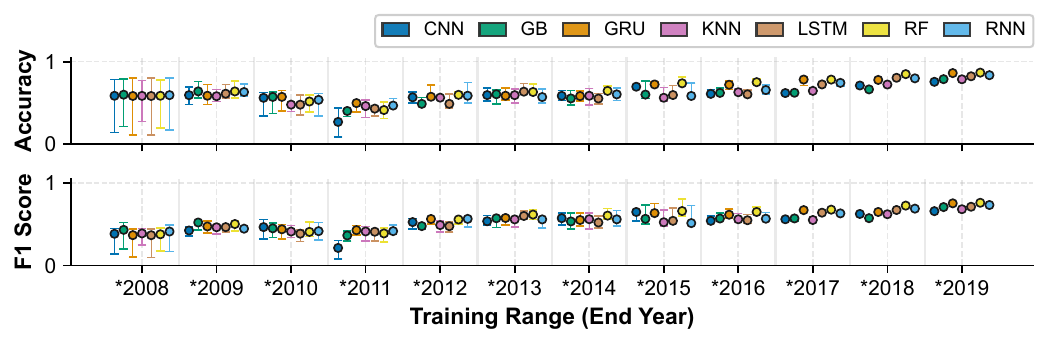}
        \caption{Emulator with dynamic features.}
        \label{fig:rq1-emu-dynamic-inc}
    \end{subfigure}

    \caption{Clean performance over time under the expanding window protocol for the emulator dataset, using static and dynamic features. For each training window that ends in year $Y_{\text{end}}$, point plots summarize accuracy and F1 on all test years after $Y_{\text{end}}$.}
    \label{fig:rq1-Incr}
\end{figure}

We analyze clean detection performance as a function of temporal separation between the training year $Y_1$ and the evaluation year $Y_2$. Figures~\ref{fig:rq1-cross} and~\ref{fig:rq1-Incr} report accuracy and macro F1 across classifiers as functions of $Y_1$, with the same-year baseline (BL) serving as the in-distribution reference. As $Y_2$ progresses beyond $Y_1$, both metrics systematically decline, indicating performance degradation associated with temporal distribution shift.

The degradation is more pronounced under the cross-year protocol, where models are deployed without updates. Models trained on earlier years (e.g., 2008--2011) experience larger performance reductions when evaluated on later $Y_2$, suggesting substantial distributional divergence over longer temporal gaps. In Figures~\ref{fig:rq1-cross}, point plot medians decrease and interquartile ranges widen as the gap increases, reflecting both reduced central performance and increased variability across classifiers.

For mid-range training years (2012--2015), degradation remains observable but is comparatively attenuated, while models trained on more recent years exhibit smaller reductions when evaluated on temporally proximate data. Overall, these results demonstrate that clean predictive performance deteriorates as a monotonic function of temporal separation, consistent with increasing non-stationarity in the underlying data distribution.

The expanding-window protocol moderates this degradation by incorporating a broader temporal span of observations into training. Relative to cross-year models, median accuracy and macro F1 at future $Y_2$ are consistently higher, and variability across classifiers is reduced (Figure~\ref{fig:rq1-Incr}). However, performance does not converge to the same-year baseline as temporal separation increases, indicating that cumulative retraining narrows but does not close the distributional gap.

Comparisons across feature spaces do not reveal a uniformly dominant representation. Static features frequently exhibit higher medians and narrower interquartile ranges, particularly on the real-device dataset, suggesting comparatively greater stability across classifiers. Dynamic features achieve comparable or lower performance depending on classifier family and training range, especially on the emulator dataset, and display greater variability across years. These differences are consistent with distinct structural (static) and behavioral (dynamic) signal characteristics and their differential sensitivity to temporal evolution.

Classifier family also influences temporal stability. Tree-based models (RF and GB) exhibit comparatively stable performance across years and generally higher medians relative to KNN, particularly for early training years (2008--2010). Neural models show improved performance when trained on more recent years, with stronger results observed in later evaluation periods (e.g., 2016--2017). This suggests that representation capacity interacts with recency of training data under non-stationary conditions.

Both emulator and real-device datasets exhibit the same overall directional trend, although real-device models attain higher absolute performance levels. Collectively, the results for RQ1 indicate that clean predictive performance degrades as temporal separation increases, and that the magnitude and stability of this degradation depend on the evaluation protocol, feature representation, and classifier family.

\subsection{RQ2: Adversarial Vulnerability Under Drift}

We examine how adversarial vulnerability evolves with increasing temporal separation between the training year $Y_1$ and the evaluation year $Y_2$. Vulnerability is quantified using Adversarial Accuracy (AA) and Attack Success Rate (ASR) under FGSM and SPSA, across both cross-year and expanding-window protocols and for static and dynamic feature spaces.

\begin{figure}[H]
    \centering
    \begin{subfigure}[t]{0.99\textwidth}
        \centering
        \includegraphics[width=\textwidth]{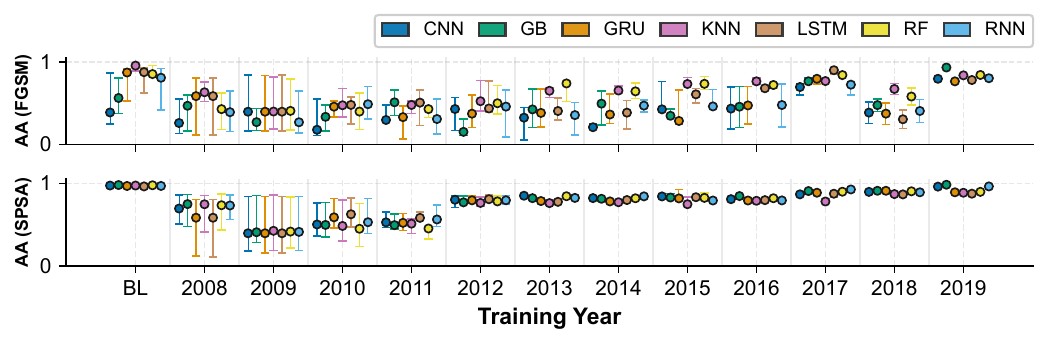}
        \caption{Emulator with static features.}
        \label{fig:rq2-emu-static-cross}
    \end{subfigure}

    \vspace{0.8em}

    \begin{subfigure}[t]{0.99\textwidth}
        \centering
        \includegraphics[width=\textwidth]{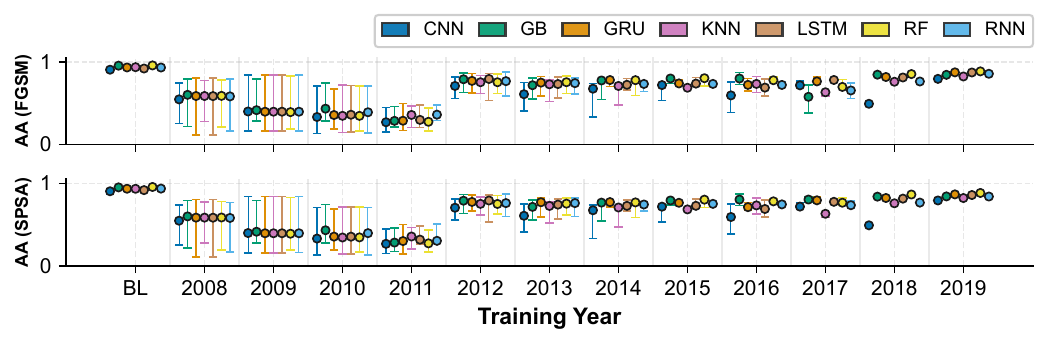}
        \caption{Emulator with dynamic features.}
        \label{fig:rq2-emu-dynamic-cross}
    \end{subfigure}

\caption{Adversarial Accuracy (AA) for cross-year evaluation on the emulator dataset using static (a) and dynamic (b) features. For each training year $Y_1$, point plots summarize AA over all future evaluation years $Y_2 > Y_1$ under FGSM and SPSA attacks. The horizontal BL line denotes the same-year baseline, where models are trained and evaluated on $Y_1$.}    \label{fig:rq2-cross-aa-emu}
\end{figure}

\BfPara{Adversarial Accuracy} AA decreases systematically as the train--test gap widens. Under the cross-year protocol (Figure~\ref{fig:rq2-cross-aa-emu}), AA declines as $Y_2$ advances relative to $Y_1$. For each training year $Y_1$, the same-year baseline (BL) consistently yields the highest AA, while evaluations on subsequent years show progressively lower medians. This pattern indicates that robustness degrades when models trained on earlier data distributions are exposed to temporally shifted inputs.

The decline is particularly pronounced for early training years (2008--2011), where median AA decreases substantially and interquartile ranges widen, reflecting both central robustness loss and increased variability across classifiers. For more recent training years, reductions in AA are comparatively attenuated over temporally proximate evaluations, suggesting that robustness erosion correlates with the magnitude of distributional shift.

Robustness differences also emerge across feature spaces and attack types. Under FGSM, models trained on dynamic features generally achieve higher AA than those trained on static features on the emulator dataset (Figures~\ref{fig:rq2-emu-dynamic-cross} and~\ref{fig:rq2-emu-static-cross}). In contrast, under SPSA, dynamic-feature models exhibit greater variability and lower AA in several configurations. This distinction is consistent with the interaction between the query-based optimization dynamics of SPSA and the higher-dimensional, continuous-valued structure of dynamic feature representations, relative to binary static indicators.

\begin{figure}[H]
    \centering

    \begin{subfigure}[t]{0.95\linewidth}
        \centering
        \includegraphics[width=\linewidth]{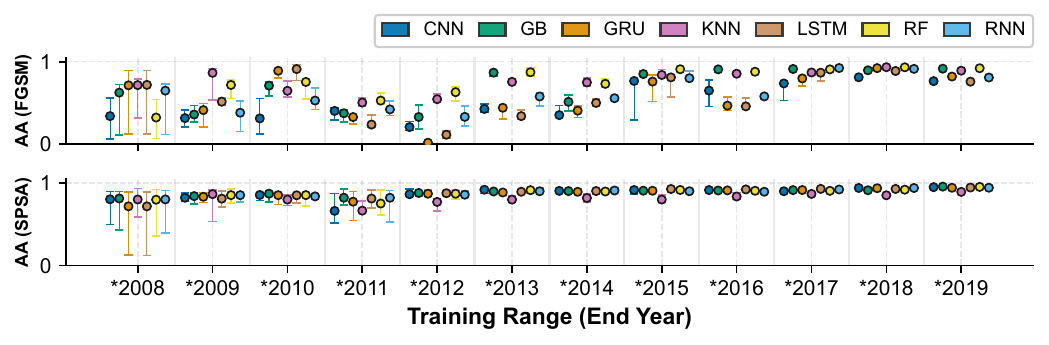}
        \caption{Real device with static features.}
        \label{fig:rq2-real-static-incr}
    \end{subfigure}

    \vspace{0.7em}

    \begin{subfigure}[t]{0.95\linewidth}
        \centering
        \includegraphics[width=\linewidth]{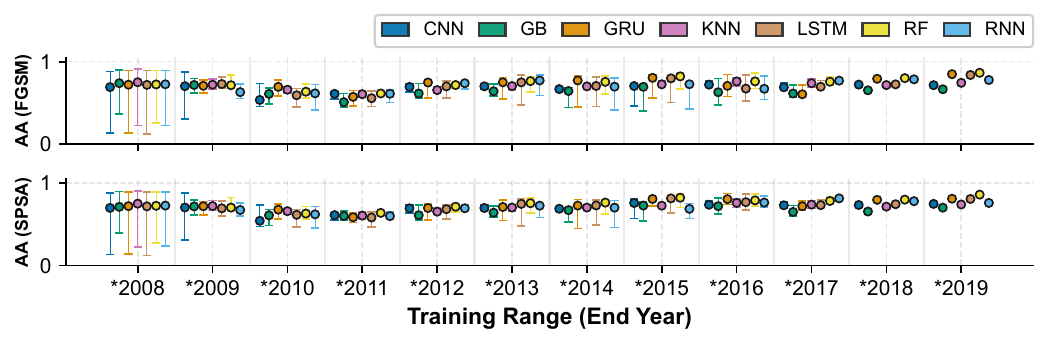}
        \caption{Real device with dynamic features.}
        \label{fig:rq2-real-dynamic-incr}
    \end{subfigure}

    \caption{Adversarial accuracy (AA) for expanding window models on the real device, using static (a) and dynamic (b) features. For each training window that starts in 2008 and ends in year $Y_{\text{end}}$, point plots aggregate AA over all corresponding test years under the FGSM and SPSA attacks. The horizontal BL line marks the reference baseline trained and evaluated on the same year.}
    \label{fig:rq2-incr-aa}
\end{figure}

Expanding-window retraining improves robustness by exposing the model to a broader temporal range, but it may also retain outdated distributions from earlier years. These older samples can stabilize learning, yet they may dilute adaptation to recent benign behavior and newly emerging malware families. This explains why expanding-window retraining reduces the robustness gap relative to cross-year deployment but does not fully restore same-year robustness. On the real-device dataset (Figure~\ref{fig:rq2-incr-aa}), AA remains closer to the same-year baseline (BL) as the training window expands, and dispersion across classifiers is reduced relative to the cross-year protocol. This indicates that cumulative exposure to temporally diverse data improves stability under adversarial perturbation. To better distinguish ordinary temporal degradation from adversarial-specific degradation, we jointly examine the clean-performance trends in Figures~3--4 and the adversarial-performance trends in Figures~5--7 across increasing temporal gaps. Under the cross-year FGSM setting, adversarial accuracy declines more sharply than the corresponding clean accuracy in several configurations, particularly for early training years and static features. This suggests that the observed reduction in adversarial accuracy is not explained solely by ordinary clean-performance degradation, but is also associated with increased sensitivity to adversarial perturbations under temporal distribution shift. However, this effect is configuration-dependent and is not consistently observed across all attacks and feature spaces.

However, for sufficiently distant evaluation years $Y_2$, AA remains consistently below the baseline, demonstrating that residual robustness loss persists despite incremental retraining. In other words, adaptive updates narrow the robustness gap but do not fully restore in-distribution adversarial performance under substantial temporal separation.

Differences across feature spaces remain observable under the expanding-window protocol. Static features exhibit comparatively stable AA across years, particularly on the real-device dataset, suggesting greater resilience to temporal variability. Dynamic features benefit from retraining but retain higher sensitivity to temporal separation, especially under SPSA (Figures~\ref{fig:rq2-real-static-incr} and~\ref{fig:rq2-real-dynamic-incr}). This pattern is consistent with the greater representational flexibility of dynamic signals, which may also render them more susceptible to distributional evolution under query-based perturbations.

Overall, these results indicate that incremental retraining attenuates, but does not fully offset, the interaction between temporal drift and adversarial vulnerability.

\BfPara{Effect of feasibility projection}
Before interpreting ASR values, we examine whether generated perturbations remain active after feasibility projection. Table~\ref{tab:projection-diagnostics} reports raw and projected perturbation magnitudes under the cross-year protocol. The FGSM retains non-zero projected perturbations across both datasets and feature spaces. Regarding static features, FGSM modifies 4.33 features on the emulator dataset and 4.86 features on the real-device dataset after projection, compared with a raw budget of nine features. For dynamic features, FGSM modified 5.16 and 4.50 features after projection on the emulator and real-device datasets, respectively.
In contrast, the SPSA is substantially weakened by feasibility projection. Its projected $L_0$ is zero in both static feature settings and remains small for dynamic features. The projected ASR of the SPSA is zero across the reported cross-year configurations. The projection-removal percentages further confirm this behavior: FGSM has negligible removal in all settings, whereas SPSA removal reaches 1.000 for both static feature settings and remains high for dynamic features. Therefore, a near-zero SPSA ASR should not be interpreted as evidence of strong model robustness. Instead, this indicates that SPSA perturbations are often removed by the imposed discrete feasibility constraints. Accordingly, we retain SPSA as an auxiliary diagnostic and use FGSM as the main constrained feature-space attack in the revised adversarial analysis.

\begin{table}[t]
\centering
\small
\caption{Perturbation effectiveness after feasibility projection under the cross-year protocol for all models. The values are the medians of all seven target classifiers. $L_0$ reports the number of modified malware features before and after the projection. Removal denotes the fraction of raw perturbation coordinates removed by projection.}
\label{tab:projection-diagnostics}
\resizebox{\linewidth}{!}{
\begin{tabular}{llrrrrrrrr}
\toprule
Dataset & Features 
& FGSM raw $L_0$ & FGSM proj. $L_0$ & FGSM removal & FGSM ASR
& SPSA raw $L_0$ & SPSA proj. $L_0$ & SPSA removal & SPSA ASR \\
\midrule
Emulator & Dynamic & 18.11 & 5.16 & 0.003 & 0.003 & 4.78 & 0.00 & 0.971 & 0.000 \\
Emulator & Static  & 9.00  & 4.33 & 0.000 & 0.222 & 9.00 & 0.00 & 1.000 & 0.000 \\
Real device & Dynamic & 18.52 & 4.50 & 0.006 & 0.002 & 6.15 & 0.57 & 0.434 & 0.000 \\
Real device & Static  & 9.00  & 4.86 & 0.000 & 0.182 & 9.00 & 0.00 & 1.000 & 0.000 \\
\bottomrule
\end{tabular}
}
\end{table}

\begin{figure}[H]
    \centering
    \begin{subfigure}[t]{0.99\textwidth}
        \centering
        \includegraphics[width=\textwidth]{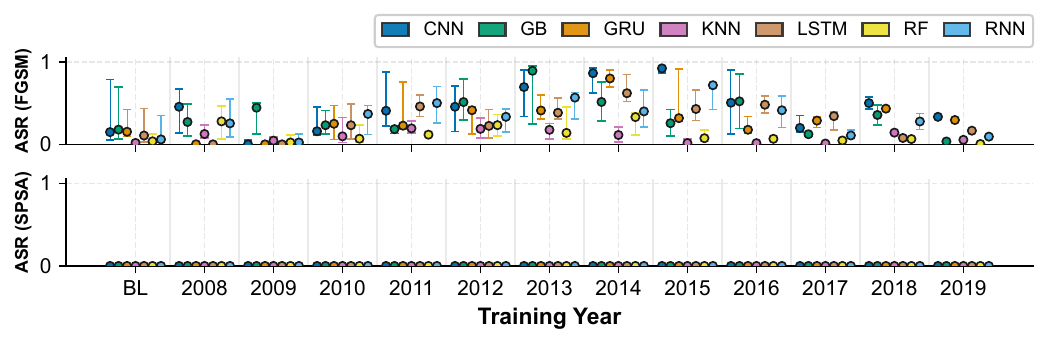}
        \caption{Real device with  static features.}
        \label{fig:rq2-asr-real-static-cross}
    \end{subfigure}

    \vspace{0.7em}

    \begin{subfigure}[t]{0.99\textwidth}
        \centering
        \includegraphics[width=\textwidth]{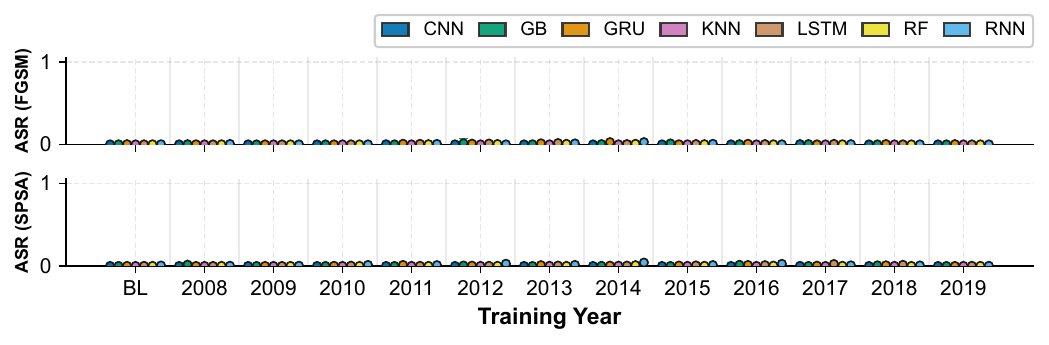}
        \caption{Real device with dynamic features.}
        \label{fig:rq2-asr-real-dynamic-cross}
    \end{subfigure}

    \caption{Attack success rate (ASR) for cross year models on the real device, using static (a) and dynamic (b) features. For each training year $Y_1$, point plots aggregate ASR over all test years $Y_2 > Y_1$ under the FGSM and SPSA attacks. Higher ASR means that adversarial perturbations are more successful.}
    \label{fig:rq2-asr-cross-real}
\end{figure}

\BfPara{Attack Success Rate} The Attack Success Rate (ASR) provides a complementary perspective on adversarial vulnerability by measuring the proportion of originally correct predictions that are flipped by perturbations. In cross-year experiments on the real-device dataset (Figure~\ref{fig:rq2-asr-cross-real}), the largest deviations from the same-year baseline occur under FGSM on static features (Figure~\ref{fig:rq2-asr-real-static-cross}). In these configurations, ASR increases for early training years and interquartile ranges widen, indicating both a higher incidence of successful perturbations and greater variability across classifiers as temporal separation grows.

In contrast, dynamic features exhibit ASR values close to zero across most years, and SPSA produces near-zero ASR for both feature spaces (Figure~\ref{fig:rq2-asr-real-dynamic-cross}). Under the imposed feasibility constraints and bounded edit budgets, SPSA rarely induces explicit label flips. As a result, its impact is more prominently reflected in reductions in Adversarial Accuracy (AA) rather than large increases in ASR. This distinction highlights that different attack mechanisms interact differently with temporal drift, depending on feature representation and perturbation structure.

Expanding-window retraining shifts AA closer to the same-year baseline relative to cross-year deployment, particularly on the real-device dataset (Figure~\ref{fig:rq2-incr-aa}). Dispersion across classifiers is also reduced. Nonetheless, for sufficiently distant evaluation years $Y_2$, AA remains below BL, reflecting persistent robustness sensitivity under temporal separation.

Overall, expanding-window retraining attenuates attack success relative to cross-year deployment, indicating that cumulative exposure to temporally diverse data improves robustness. However, residual vulnerability persists under FGSM for specific model--feature combinations, demonstrating that incremental updates do not fully neutralize adversarial susceptibility under sustained distributional evolution.

\BfPara{Summary} Increasing temporal separation is associated with elevated adversarial vulnerability, reflected primarily by declining AA under cross-year evaluation and, in selected configurations, higher ASR for early training years. Expanding-window retraining narrows this gap relative to the same-year baseline but does not eliminate it under continued distributional shift. Residual vulnerability is most pronounced under FGSM, particularly for static features, whereas SPSA and dynamic-feature ASR remain near zero except for isolated outliers. Collectively, these results indicate that temporal separation is associated with disproportionate adversarial degradation in selected configurations, while adaptive retraining provides partial but incomplete mitigation.

\begin{figure}[H]
    \centering
    \begin{subfigure}[t]{0.99\textwidth}
        \centering
        \includegraphics[width=\textwidth]{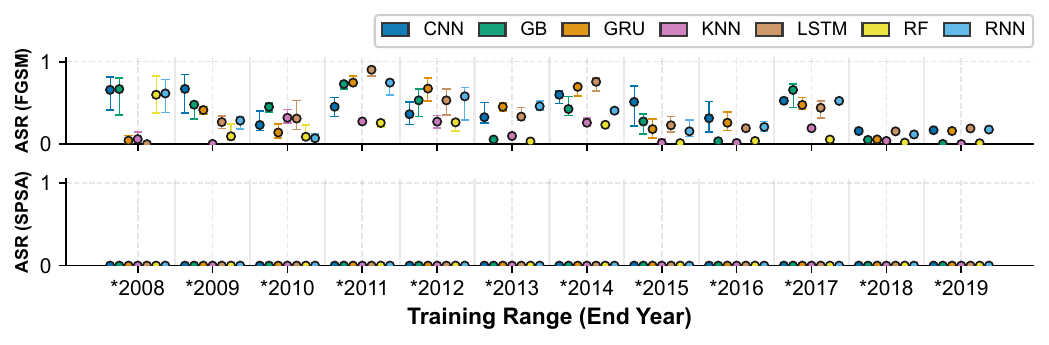}
        \caption{Emulator dataset with static features.}
        \label{fig:rq2-asr-emu-static-incr}
    \end{subfigure}

    \vspace{0.7em}

    \begin{subfigure}[t]{0.99\textwidth}
        \centering
        \includegraphics[width=\textwidth]{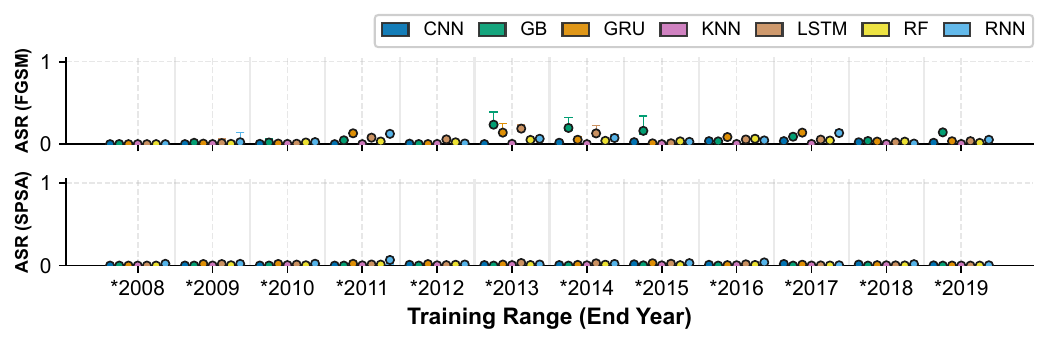}
        \caption{Emulator dataset with dynamic features.}
        \label{fig:rq2-asr-emu-dynamic-incr}
    \end{subfigure}

    \caption{Attack success rate (ASR) for expanding window (incremental) models on the emulator device, using static (a) and dynamic (b) features. For each training window end year, point plots aggregate ASR over all corresponding test years under the FGSM and SPSA attacks. Lower ASR than in the cross year case shows that incremental retraining mitigates drift induced adversarial vulnerability.}
    \label{fig:rq2-asr-incr-emu}
\end{figure}

\subsection{RQ3: Quantifying Drift and Adversarial Vulnerability}

\begin{figure}[H]
    \centering
    \begin{subfigure}[t]{0.95\linewidth}
        \centering
        \includegraphics[width=\linewidth]{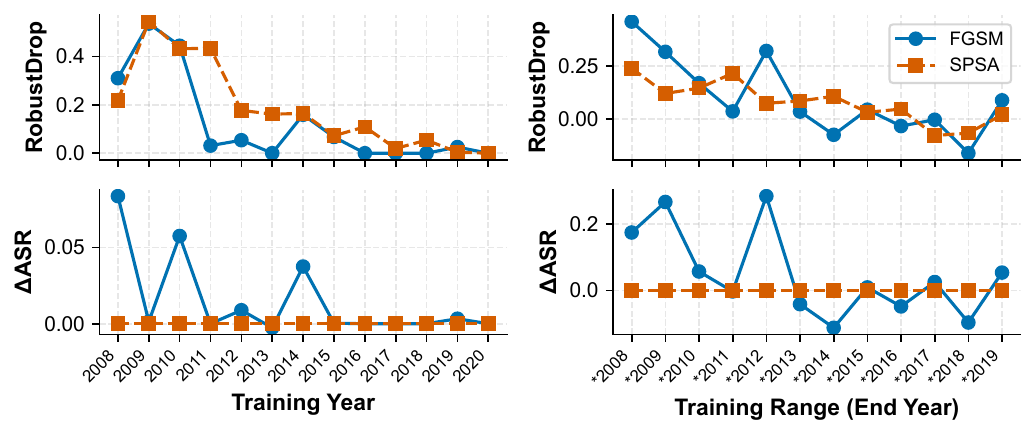}
        \caption{Emulator dataset with static features.}
        \label{fig:rq3-emu-static}
    \end{subfigure}

    \vspace{0.7em}

    \begin{subfigure}[t]{0.95\linewidth}
        \centering
        \includegraphics[width=\linewidth]{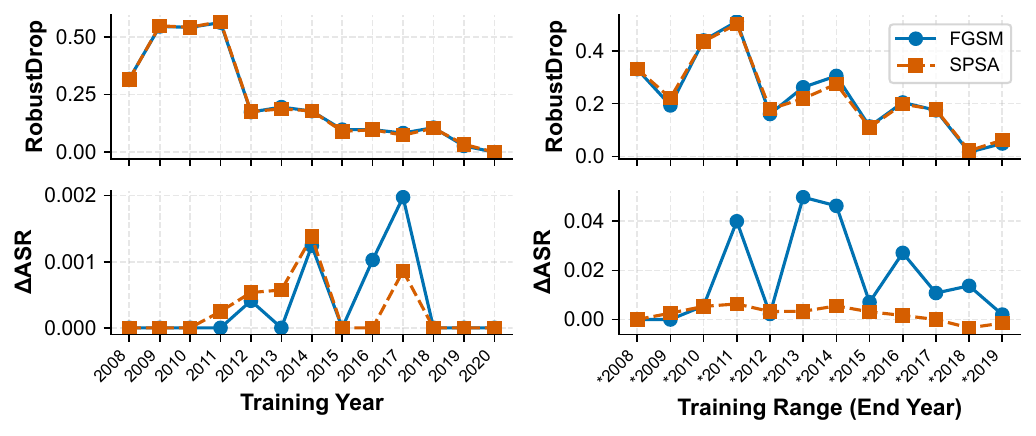}
        \caption{Emulator dataset with dynamic features.}
        \label{fig:rq3-emu-dynamic}
    \end{subfigure}

    \caption{Linkage metrics for the emulator device. Each 2$\times$2 panel uses the training year or training window end year and shows the median RobustDrop (top row) and $\Delta$ASR (bottom row) for FGSM and SPSA, aggregated over $(Y_1, Y_2)$ pairs.}
    \label{fig:rq3-emu}
\end{figure}

\begin{figure}[H]
    \centering
    \begin{subfigure}[t]{0.95\linewidth}
        \centering
        \includegraphics[width=\linewidth]{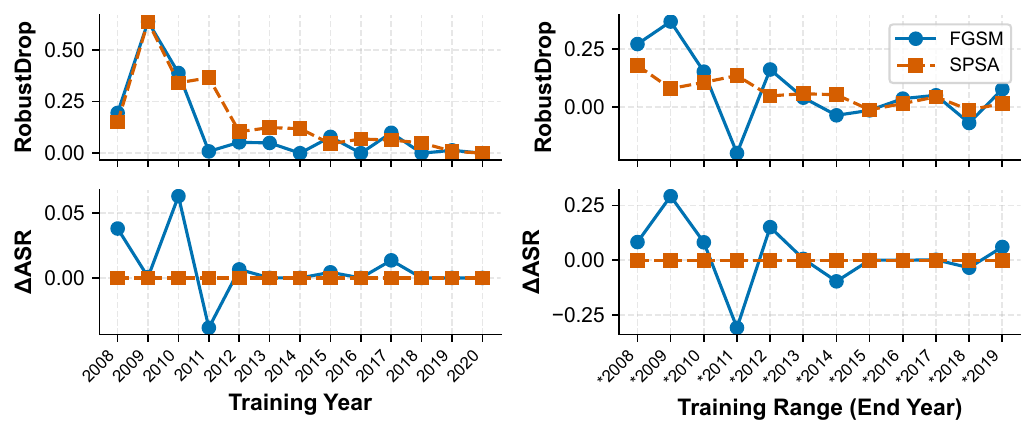}
        \caption{Real device with static features.}
        \label{fig:rq3-real-static}
    \end{subfigure}

    \vspace{0.7em}

    \begin{subfigure}[t]{0.95\linewidth}
        \centering
        \includegraphics[width=\linewidth]{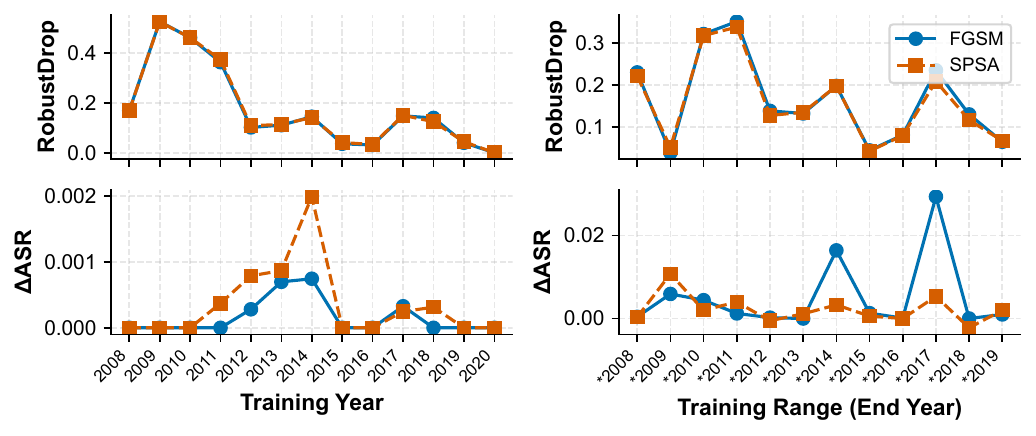}
        \caption{Real device with dynamic features.}
        \label{fig:rq3-real-dynamic}
    \end{subfigure}

    \caption{Linkage metrics for the real device. The panels summarize how RobustDrop, $\Delta\text{ASR}$, and AAF evolve as the training data move forward in time, under both FGSM and SPSA attacks.}
    \label{fig:rq3-real}
\end{figure}

This research question examines the association between temporal drift and adversarial vulnerability, and the extent to which expanding-window retraining moderates this relationship. The linkage metrics are used to express robustness changes as temporal contrasts relative to the same-year baseline, rather than as replacements for AA and ASR. For each $(Y_1,Y_2)$ pair, we compute \ding{172} RobustDrop (RD), defined as the reduction in adversarial accuracy relative to the same-year baseline; \ding{173} $\Delta\mathrm{ASR}$, the change in attack success rate relative to baseline; and \ding{174} AAF, the ratio of attack success in $Y_2$ to its baseline value. Figures~\ref{fig:rq3-emu} and~\ref{fig:rq3-real} report the RD (top rows) and $\Delta\mathrm{ASR}$ (bottom rows) across years under the cross-year and expanding-window protocols. Table~\ref{tab:rq3-summary-fgsm-spsa} summarizes Spearman correlations and median AAF values.

To support the temporal-gap analysis, we compute bootstrap 95\% confidence intervals for median metrics by $\Delta T$ and Spearman correlation coefficients with corresponding $p$-values between $\Delta T$ and each robustness metric. We report the cross-year RobustDrop association in the main text because it directly captures non-adaptive deployment under increasing temporal separation. The results show a positive association between the temporal gap and FGSM RobustDrop across both datasets and feature spaces. This indicates that a larger train--test temporal separation is associated with a greater adversarial accuracy reduction under the cross-year deployment protocol. We use RobustDrop for this summary because it remains interpretable when ASR is low, unlike ratio-based measures such as AAF.

\begin{table}[t]
\centering
\small
\caption{Spearman association between temporal gap $\Delta T$ and FGSM RobustDrop under the cross-year protocol. RobustDrop measures the reduction in adversarial accuracy relative to the same-year baseline.}
\label{tab:spearman-robustdrop}
\begin{tabular}{llcc}
\toprule
Dataset & Features & Spearman $\rho$ & $p$-value \\
\midrule
\multirow{2}{*}{Emulator}
    & Dynamic & 0.681 & 0.010 \\
    & Static  & 0.758 & 0.003 \\
\midrule
\multirow{2}{*}{Real device}
    & Dynamic & 0.560 & 0.046 \\
    & Static  & 0.692 & 0.009 \\
\bottomrule
\end{tabular}
\end{table}

RD exhibits a consistent temporal pattern across devices and feature spaces. Under cross-year evaluation (left columns of Figures~\ref{fig:rq3-emu} and~\ref{fig:rq3-real}), RD is largest for early training years and decreases as $Y_1$ increases, indicating greater reductions in adversarial accuracy for models trained on temporally distant data. Under expanding-window retraining (right columns), RD values are consistently smaller and vary more smoothly across years, indicating partial mitigation of robustness loss through cumulative retraining.

Table~\ref{tab:rq3-summary-fgsm-spsa} quantifies this trend: $\rho_X(\mathrm{RD})<0$ across all evaluated settings, demonstrating a negative association between training year and robustness degradation. The magnitude of this association is stronger for static features under expanding-window retraining (e.g., emulator static features: $\rho_X(\mathrm{RD})=-0.562$ for FGSM and $-0.531$ for SPSA), consistent with the visual patterns in the corresponding figures.

$\Delta\mathrm{ASR}$ provides complementary information and varies across attack types and feature spaces. For static features, FGSM exhibits the largest early-year deviations from zero (Figures~\ref{fig:rq3-emu-static} and~\ref{fig:rq3-real-static}, bottom-left panels), reflecting greater increases in attack-induced label flips for models trained on earlier years. Under expanding-window retraining (bottom-right panels), these deviations are reduced, consistent with attenuation of vulnerability relative to cross-year deployment.

For SPSA, $\Delta\mathrm{ASR}$ remains close to zero in multiple configurations, particularly for static features. In such cases, correlations are undefined (``--'') in Table~\ref{tab:rq3-summary-fgsm-spsa}, as limited variation in ASR precludes meaningful rank association. For dynamic features, $\Delta\mathrm{ASR}$ values are generally small and characterized by localized peaks rather than monotonic trends (Figures~\ref{fig:rq3-emu-dynamic} and~\ref{fig:rq3-real-dynamic}), suggesting year- or model-specific effects rather than a consistent temporal pattern. One contributing factor is projection to a discrete feasible domain, where quantized updates may limit boundary crossings and reduce systematic amplification of attack success over time.

The Adversarial Amplification Factor (AAF) evaluates whether attack success in $Y_2$ exceeds its same-year baseline. Table~\ref{tab:rq3-summary-fgsm-spsa} reports AAF values greater than 1 in several expanding-window configurations, particularly for dynamic features, indicating amplification relative to baseline. Because AAF is normalized by baseline ASR, its magnitude is sensitive when baseline ASR is near zero. Accordingly, AAF is interpreted jointly with RD and $\Delta\mathrm{ASR}$ rather than in isolation.

\BfPara{Interpretation of Drift–Robustness Coupling}
One plausible mechanism underlying the observed association is geometric. Temporal drift alters feature distributions and may reshape the effective decision boundary learned at $Y_1$. When evaluated on future data $Y_2$, samples may lie closer to this boundary due to distributional shift, even if clean classification remains correct. Reduced margin or altered gradient alignment can increase the sensitivity of predictions to small perturbations, thereby amplifying adversarial vulnerability beyond what is explained by clean accuracy degradation alone. Under expanding-window retraining, the boundary adapts to a broader temporal span, which stabilizes margins and partially mitigates this effect, but continued evolution may still induce boundary misalignment over time.

\BfPara{Effect of control baselines}
Table~\ref{tab:protocol-comparison} compares cross-year deployment with full expanding-window, size-matched expanding-window, and recent-window retraining. The results show that retraining effects are configuration-dependent. Static features generally benefit from expanding-window variants, while dynamic features do not always improve under cumulative historical training. The size-matched baseline further shows that the observed changes cannot be explained by the training-set size alone. Overall, the added baselines indicate that robustness under temporal drift depends on temporal coverage, recency, and feature representation.

\begin{table*}[t]
\centering
\small
\caption{Protocol comparison across temporal deployment strategies. Cross-year trains on a single year and tests on future years. Expanding Window (EW)-full trains on all historical data up to $Y_1$. EW-size-matched samples from the same historical pool while matching the single-year training size. Recent-window trains on the most recent three years up to $Y_1$. Values are the medians over all seven classifiers.}
\label{tab:protocol-comparison}
\resizebox{\textwidth}{!}{
\begin{tabular}{lllrrrrr}
\toprule
Dataset & Features & Protocol & Clean Acc. & FGSM AA & FGSM ASR & RobustDrop & Train size \\
\midrule
\multirow{8}{*}{Emulator}
& \multirow{4}{*}{Dynamic}
& Cross-year & 0.743 & 0.728 & 0.003 & 0.180 & 2,220 \\
& & EW-full & 0.605 & 0.572 & 0.031 & 0.227 & 23,746 \\
& & EW-size-matched & 0.539 & 0.491 & 0.026 & 0.234 & 3,668 \\
& & Recent-window & 0.634 & 0.604 & 0.029 & 0.202 & 8,414 \\
\cmidrule(lr){2-8}
& \multirow{4}{*}{Static}
& Cross-year & 0.807 & 0.498 & 0.222 & 0.105 & 2,220 \\
& & EW-full & 0.827 & 0.563 & 0.256 & 0.036 & 23,746 \\
& & EW-size-matched & 0.805 & 0.539 & 0.288 & 0.068 & 3,668 \\
& & Recent-window & 0.807 & 0.508 & 0.298 & 0.105 & 8,414 \\
\midrule
\multirow{8}{*}{Real device}
& \multirow{4}{*}{Dynamic}
& Cross-year & 0.782 & 0.773 & 0.002 & 0.169 & 3,747 \\
& & EW-full & 0.693 & 0.666 & 0.026 & 0.187 & 25,386 \\
& & EW-size-matched & 0.651 & 0.626 & 0.023 & 0.185 & 3,747 \\
& & Recent-window & 0.688 & 0.675 & 0.015 & 0.175 & 12,646 \\
\cmidrule(lr){2-8}
& \multirow{4}{*}{Static}
& Cross-year & 0.861 & 0.596 & 0.182 & 0.096 & 3,747 \\
& & EW-full & 0.874 & 0.615 & 0.236 & 0.014 & 25,386 \\
& & EW-size-matched & 0.865 & 0.642 & 0.188 & 0.036 & 3,747 \\
& & Recent-window & 0.860 & 0.530 & 0.278 & 0.068 & 12,646 \\
\bottomrule
\end{tabular}
}
\end{table*}

\BfPara{Summary} The linkage metrics collectively indicate a measurable association between temporal separation and adversarial vulnerability under the evaluated feature-space setting. RobustDrop consistently captures this relationship across devices, feature spaces, and evaluation protocols, showing larger reductions in adversarial accuracy for earlier training years and partial mitigation under incremental retraining. In contrast, $\Delta\mathrm{ASR}$ and AAF are most informative when ASR is non-negligible, and are interpreted cautiously for SPSA configurations where ASR is near-zero or constant. Overall, the evidence supports a temporal association with adversarial-accuracy degradation, while ASR amplification remains configuration-dependent.

In contrast, $\Delta\mathrm{ASR}$ and AAF vary across attack types and feature representations. Amplification relative to baseline is most visible under FGSM, particularly in settings where attack-induced label flips are non-trivial. When ASR remains near zero due to discrete validity constraints, $\Delta\mathrm{ASR}$ and AAF exhibit limited variation. These findings indicate that the strength and form of the drift–robustness linkage depend on both the attack mechanism and the structure of the feasible feature domain.

\begin{table}[H]
\centering
\small
\caption{Linkage summary (Spearman). (CY) cross-year, (EW) expanding-window, (Dyn) dynamic, (Sta) static. $X$ denotes the training year for CY and EW). Each entry reports FGSM / SPSA and is aggregated over all $(Y_1,Y_2)$ pairs and all seven target models. ``--'' indicates undefined correlation due to near-constant $\Delta$ASR.}
\label{tab:rq3-summary-fgsm-spsa}
\scalebox{0.75}{
\begin{tabular}{lllccc}
\toprule
Dataset & Features & Prot. &
$\rho_X(\text{RD})$ & $\rho_X(\Delta\text{ASR})$ & AAF \\
\cmidrule(lr){4-6}
 &  &  & FGSM / SPSA & FGSM / SPSA & FGSM / SPSA \\
\midrule
\multirow{4}{*}{Emulator}
& \multirow{2}{*}{Dyn} & CY & -0.360 / -0.369 &  0.076 /  0.014 & 0.223 / 0.164 \\
&                      & EW & -0.287 / -0.339 &  0.347 / -0.039 & 1.553 / 1.610 \\
\cmidrule{2-6}
& \multirow{2}{*}{Sta} & CY & -0.461 / -0.432 & -0.246 / --     & 1.015 / 0 \\
&                      & EW & -0.562 / -0.531 & -0.415 / --     & 1.812 / 0 \\
\midrule
\multirow{4}{*}{Real device}
& \multirow{2}{*}{Dyn} & CY & -0.342 / -0.347 & -0.022 /  0.033 & 0.169 / 0.436 \\
&                      & EW & -0.126 / -0.170 & -0.051 / -0.195 & 1.109 / 1.170 \\
\cmidrule{2-6}
& \multirow{2}{*}{Sta} & CY & -0.415 / -0.426 & -0.195 / --     & 0.999 / 0 \\
&                      & EW & -0.469 / -0.444 & -0.339 / --     & 1.314 / 0 \\
\bottomrule
\end{tabular}
}
\end{table}

We note that the reported trends should be interpreted as associations observed under the evaluated transfer-based feature-space setting rather than as causal evidence that temporal drift directly increases adversarial vulnerability across all deployment conditions.

\section{Threats to Validity}\label{sec:Threats}

\BfPara{Internal validity} Adversarial evaluation was conducted in feature space under explicit feasibility constraints. Each attack modifies at most $k_{\text{sta}}=\lceil 0.05\, d_{\text{sta}}\rceil$ and $k_{\text{dyn}}=\lceil 0.05\, d_{\text{dyn}}\rceil$ features, with fixed perturbation strengths ($\epsilon_{\text{sta}}=0.55$, $\epsilon_{\text{dyn}}=0.20$). Dynamic perturbations are generated in standardized space and subsequently mapped back to count-based representations. After inverse scaling, rounding, and enforcing a per-feature cap of one count change, many continuous perturbations may be removed or reduced to $0$ or $\pm 1$ count changes. 
To make this limitation explicit, we report projection diagnostics before and after enforcing feature-domain constraints. These diagnostics show that FGSM generally remains active after projection, whereas SPSA is often substantially weakened because many of its raw perturbation coordinates are removed by the projection step. Therefore, near-zero SPSA ASR is not interpreted as evidence of strong model robustness. Instead, SPSA is treated as an auxiliary constrained diagnostic, and the main adversarial analysis relies primarily on FGSM and projected adversarial accuracy.

These constraints preserve representation level validity and ensure that adversarial samples remain within the  feasible domain of feature vector. However, they also restrict attack flexibility and may limit decision-boundary crossings. As a result, ASR may remain low in certain configurations—particularly for SPSA and dynamic features—even when AA decreases. More permissive edit budgets, alternative feasibility projections, or problem-space manipulations could produce different attack effectiveness. Accordingly, our conclusions apply within the defined feature-space threat model.

\BfPara{Problem-space realizability}
Our adversarial evaluation is conducted in feature space and does not construct modified APKs. Although the projection step ensures that perturbed vectors satisfy representation-level constraints, such as binary static features and non-negative integer dynamic counts, this does not guarantee problem-space realizability. In particular, a modified permission, intent, or system-call-count vector may not correspond to a functional, installable, and behavior-preserving Android application. Therefore, the reported adversarial results should be interpreted as feature-space robustness evidence, not as direct measurement of real-world APK evasion success.

\BfPara{External validity} Our analysis is based on static and dynamic feature representations extracted from emulator and real-device executions within the KronoDroid dataset. Alternative datasets, feature extraction pipelines, platform versions, or data collection procedures may exhibit different temporal dynamics and distributional shifts. 

Similarly, the relative behavior of FGSM and SPSA may vary under different feature encodings or attack implementations. We evaluated a fixed set of target models and employed a logistic regression surrogate for transfer-based attacks. Different model families, surrogate architectures, or stronger adaptive attackers could alter quantitative vulnerability levels. Therefore, the reported findings should be interpreted as representative of the evaluated configurations rather than universally generalizable across all Android malware detection systems.

\BfPara{Construct validity} Robustness is quantified using AA, ASR, and the linkage metrics RobustDrop, $\Delta$ASR, and AAF. These metrics capture complementary aspects of adversarial vulnerability but are sensitive to baseline behavior and attack constraints. In particular, AAF is normalized by baseline ASR and can become unstable when baseline ASR approaches zero. For this reason, AAF is interpreted jointly with RobustDrop and $\Delta$ASR rather than independently.

More broadly, the measured association between temporal drift and adversarial vulnerability depends on the operationalization of drift through temporal separation and on the selected robustness metrics. Alternative drift characterizations or robustness definitions may yield different quantitative magnitudes, although the qualitative trends observed here are consistent across devices, feature spaces, and evaluation protocols.
% -------------------------------------------------
\section{Conclusion}\label{sec:conclusion}

{We presented a longitudinal evaluation of temporal drift and adversarial robustness in Android malware detection using the KronoDroid dataset (2008--2020), with the objective of characterizing their association under transfer-based feature-space adversarial evaluation.
Across cross-year and expanding-window settings, increasing temporal separation between training and evaluation data is consistently associated with declines in clean predictive performance and reductions in adversarial accuracy relative to same-year baselines. In selected configurations, attack success rates also increase for models trained on earlier years, indicating amplified vulnerability under sustained distributional evolution.

Expanding-window retraining mitigates these effects relative to cross-year deployment by increasing temporal coverage. However, robustness does not fully recover under continued temporal separation, likely because cumulative training retains older distributions, whereas newly emerging malware and benign behaviors may not be fully represented by historical static and dynamic features. Thus, cumulative retraining narrows, but does not eliminate, robustness loss within the evaluated feature-space threat model.

The results indicate a measurable association between temporal separation and reduced adversarial accuracy in the evaluated Android malware detection setting.
More broadly, they underscore the importance of explicitly accounting for non-stationary data distributions when evaluating the robustness of intelligent detection systems. These findings motivate future work on drift monitoring mechanisms, adaptive retraining strategies, and robustness-aware temporal learning frameworks designed for evolving, adversarial environments.

Beyond the Android setting, these findings highlight a broader systems-level implication: robustness in intelligent detection systems is inherently time-dependent. Evaluations conducted under stationary assumptions may overestimate long-term reliability when models are deployed in evolving environments. As a result, adversarial robustness should be assessed not only against perturbation strength but also against temporal separation between training and deployment data. Incorporating drift-aware evaluation protocols into robustness assessment frameworks may therefore be necessary for realistic appraisal of intelligent security systems operating in non-stationary, adversarial domains. These conclusions are limited to the evaluated Android dataset family, static and dynamic KronoDroid feature representations, selected classifier families, and constrained transfer-based feature-space attacks.

\end{document}